\pdfoutput=1
\documentclass[11pt,twoside,a4paper,cmspaper,final,collab]{cms-tdr}
\def\svnVersion{882c69f-D}\def\svnDate{2025/08/30}\def\cmsCernNoTag{CERN-EP-2023-294}\def\cmsCernDate{\today}\def\cmsMessage{Published in Nature Physics as \href{http://dx.doi.org/10.1038/s41467-025-56200-6}{\doi{10.1038/s41467-025-56200-6}.}}
\begin{document}\cmsNoteHeader{HIN-20-002}

\newcommand\fz      {\ensuremath{\PfDzP{980}}\xspace}
\newcommand\nq      {\ensuremath{n_{\PQq}}\xspace}
\newcommand\vsub    {\ensuremath{v_2^{\text{sub}}}\xspace}
\newcommand\vn      {\ensuremath{v_{n}}\xspace}
\newcommand\vtwo    {\ensuremath{v_{2}}\xspace}
\newcommand\vnq     {\ensuremath{v_{n,\PQq}}\xspace}
\newcommand\vtwoq   {\ensuremath{v_{2,\PQq}}\xspace}
\newcommand\etnq    {\ensuremath{KE_{\text T}/n_{\PQq}}\xspace}
\newcommand\ptmean  {\ensuremath{\langle\pt\rangle}\xspace}
\newcommand\Ntrk    {\ensuremath{N_{\text{trk}}}\xspace}
\newcommand\nrange  {\ensuremath{185 \le \Ntrk < 250}\xspace}
\newcommand\dphi    {\ensuremath{\phi-\psi_{2}}\xspace}
\newcommand\vsubnq  {\ensuremath{\vsub/\nq}\xspace}
\newcommand\ppb     {\ensuremath{\Pp\mathrm{Pb}}\xspace}
\newcommand{\Xm}    {\HepParticleResonance{X}{3872}{}{}\Xspace}
\ifthenelse{\boolean{cms@external}}{\providecommand{\figonewidth}{0.49\textwidth}}{\providecommand{\figonewidth}{0.8\textwidth}} 
\newenvironment{dataavailability}{\bmhead{Data availability}\bgroup\sloppy}{\par\egroup}\newenvironment{codeavailability}{\bmhead{Code availability}\bgroup\sloppy}{\par\egroup}
\newenvironment{interests}{\bmhead{Competing interests}\bgroup\sloppy}{\par\egroup}
\newenvironment{authorcontributions}{\bmhead{Competing interests}\bgroup\sloppy}{\par\egroup}

\cmsNoteHeader{HIN-20-002}
\title{Elliptic anisotropy measurement of the \texorpdfstring{\fz}{f0(980)} hadron in proton-lead collisions and evidence for its quark-antiquark composition}

\author*[cern]{CMS Collaboration}

\date{\today}

\abstract{
Despite the \fz hadron having been discovered half a century ago, the question about its quark content has not been settled: 
it might be an ordinary quark-antiquark ($\PQq\PAQq$) meson, a tetraquark ($\PQq\PAQq\PQq\PAQq$) exotic state, a kaon-antikaon ($\PK\PAK$) molecule, 
or a quark-antiquark-gluon ($\PQq\PAQq\Pg$) hybrid. This paper reports strong evidence that the \fz state is an ordinary $\PQq\PAQq$ meson, 
inferred from the scaling of elliptic anisotropies (\vtwo) with the number of constituent quarks (\nq), 
as empirically established using conventional hadrons in relativistic heavy ion collisions. 
The \fz state is reconstructed via its dominant decay channel $\fz \to \PGpp\PGpm$, in proton-lead collisions recorded by the CMS experiment at the LHC, 
and its \vtwo is measured as a function of transverse momentum (\pt). It is found that the $\nq = 2$ ($\PQq\PAQq$ state) hypothesis is favored 
over $\nq = 4$ ($\PQq \PAQq \PQq \PAQq$ or $\PK\PAK$ states) by 7.7, 6.3, or 3.1 standard deviations in the $\pt <10$, 8, or 6\GeVc ranges, respectively, 
and over $\nq=3$ ($\PQq\PAQq\Pg$ hybrid state) by 3.5 standard deviations in the $\pt<8\GeVc$ range. This result represents the first determination 
of the quark content of the \fz state, made possible by using a novel approach, and paves the way for similar studies of other exotic hadron candidates.
}

\hypersetup{
pdfauthor={CMS Collaboration},%
pdftitle={Elliptic anisotropy measurement of the f0(980) hadron in proton-lead collisions and evidence of its quark-antiquark composition},%
pdfsubject={CMS},%
pdfkeywords={CMS, f0(980), exotic hadron, quark-antiquark states, mesons}}

\maketitle 

\section{Introduction}
One of the most intriguing puzzles in quantum chromodynamics (QCD), the theory describing the strong interaction, is the phenomenon of confinement. 
Confinement is the peculiar feature of the QCD color charges that they cannot be separated and are fatefully confined in 
color-neutral bound states known as hadrons. The mechanism for the color confinement is still not well understood. 
A hadron can be the usual quark-antiquark ($\PQq\PAQq$) meson or three-quark ($\PQq\PQq\PQq$) baryon, but it has been suggested 
that there also exist less conventional, ``exotic'' forms, such as tetraquarks or meson molecules ($\PQq\PAQq\PQq\PAQq$), 
pentaquarks ($\PQq\PAQq\PQq\PQq\PQq$), and dibaryons ($\PQq\PQq\PQq\PQq\PQq\PQq$)~\cite{Gell-Mann:1964ewy,Zweig:1964jf,Jaffe:1976yi}, 
where \PQq stands for a constituent quark of any flavor. Studies of these exotic states can significantly advance our understanding of how partons 
can form bound states and in which configurations. This knowledge is fundamental for a deeper understanding of QCD, 
especially in its nonperturbative regime~\cite{Jaffe:2004ph, Briceno:2017qmb}. 

Exotic hadrons are expected to be short-lived and decay into ordinary hadrons, making it challenging to decipher their original parton structure. 
The first evidence of a tetraquark or a molecular state, \Xm, was reported by the Belle experiment at KEK~\cite{PhysRevLett.91.262001}. 
Experiments at the CERN LHC, particularly LHCb, have recently observed several new candidates for tetraquarks and pentaquarks, 
as discussed, \eg, in Ref.~\cite{LHCb:2021vvq}. Those candidates all involve heavy quarks, which implies that their properties can be 
calculated in nonrelativistic QCD~\cite{Esposito:2016noz,Olsen:2017bmm}.

However, similar calculations are hard to perform for hadrons built of only light quarks, as one has to use relativistic QCD in its nonperturbative regime. 
In particular, the \fz hadron, discovered 50 years ago~\cite{Protopopescu:1973sh, Hyams:1973zf, Grayer:1974cr}, has been hypothesized to be 
an ordinary ${\PQq\PAQq}$ meson, a tetraquark state, a ${\PK\PAK}$ molecule, or a ${\PQq\PAQq}$-gluon 
hybrid state~\cite{Jaffe:1976ig, Weinstein:1990gu, Amsler:2004ps, Bugg:2004xu, Klempt:2007cp,Deandrea:2000yc}. 
The suggestion that the \fz hadron can be a tetraquark extends beyond the ground-state constituent quark model, and studies of 
such states would impact our understanding of QCD and color confinement. Despite a multitude of experimental and theoretical work, 
the nature of the \fz state has not yet been established, as is evident from Ref.~\cite{ParticleDataGroup:2024cfk} and references therein. 

This is where high-energy nuclear collisions may come to the rescue. The collisions of lead-lead (PbPb) ions at the LHC aim to recreate 
the quark-gluon plasma (QGP), widely believed to be the state of matter prevailing in the early universe, when the temperature and 
energy density were too high to allow for the formation of hadrons. They offer a universal laboratory to study various aspects of QCD, 
such as the formation of hadrons from the QGP hadronization. A large number of hadron species, presumably including exotic ones, 
are abundantly produced during and following the phase transition from the QGP to hadronic matter. 
Indeed, evidence for the production of the \Xm exotic state in PbPb collisions was reported by the CMS experiment~\cite{CMS:2021znk}. 
The QGP phase transition (hadronization), intimately connected to color confinement, is being extensively studied, 
both experimentally and theoretically. A viable way to describe hadronization is via the coalescence of quarks, 
now dressed with gluons over the phase transition, into hadrons. The coalescence model was initially proposed to describe 
the formation of deuterons in targets exposed to proton beams~\cite{Butler:1963pp} and is now commonly used to model hadronization 
in relativistic nuclear collisions~\cite{Dover:1991zn, Fries:2008hs,Hwa:2002tu,Fries:2003vb,Greco:2003xt}.

In heavy ion collisions, the azimuthal distribution of produced particles is anisotropic. The anisotropy is believed to result from 
the interactions among quarks and gluons created in these collisions, converting the initial approximately elliptical (``almond-like'') 
overlap region of the colliding nuclei with a nonzero impact parameter into the anisotropy of particle momenta~\cite{Ollitrault:1992bk}. 
It is noteworthy that the collision geometry anisotropy is generic and also present in head-on heavy ion collisions, 
as well as in proton-proton ($\Pp\Pp$) and proton-nucleus collisions, arising from fluctuations in the distribution of constituents inside 
the colliding objects~\cite{PhysRevC.77.014906}. While it initially came as a surprise when momentum anisotropy was first observed in 
$\Pp\Pp$~\cite{CMS:2010ifv,ATLAS:2015hzw,CMS:2015fgy,CMS:2016fnw} and 
proton-lead (\ppb)~\cite{CMS:2012qk,ALICE:2012eyl,ATLAS:2012cix,CMS:2013jlh,ATLAS:2014qaj,CMS:2014und,CMS:2015yux,LHCb:2015coe} collisions, 
it is by now a well-established fact. This momentum anisotropy of quarks is then inherited by the formed hadrons, thus providing information 
that can be used to experimentally determine the quark content of the hadrons~\cite{Gu:2019oyz}, as explained below. 
Since the anisotropy has been established at the LHC energies in PbPb, \ppb, and even $\Pp\Pp$ collisions with 
high multiplicity of particles produced, any of these colliding systems can be used for this type of measurements.

Azimuthal distributions of particles are often described by a Fourier series~\cite{Poskanzer:1998yz},
\begin{equation}\label{equ:flows}
	\frac{\rd N}{\rd\phi}\propto 1+\sum_{n=1}^{\infty}2\vn\cos[n(\phi-\psi_{n})],
\end{equation}
where $\phi$ is the azimuthal angle of the particle momentum vector and $\psi_{n}$ is the azimuthal angle of the $n$th harmonic plane, 
defined in each event such that ${\sum_i} \sin[ n( \phi_i-\psi_n)]=0$, where the index $i$ runs over 
all particles in an event. Details on the reconstruction of the harmonic planes using event observables are given in the Methods section. 
The coefficients \vn, called anisotropic flow parameters, generally depend on the particle transverse momentum (\pt) and rapidity ($y$). 
The \vtwo coefficient, called the elliptic flow, describes the dominant anisotropic component. The second-order harmonic plane angle $\psi_2$ 
is an approximation of the azimuthal angle of the reaction plane, which is defined by the line connecting the centers of the colliding nuclei and the beam line.
\begin{figure}[tb]
\centering
\includegraphics[width=\figonewidth]{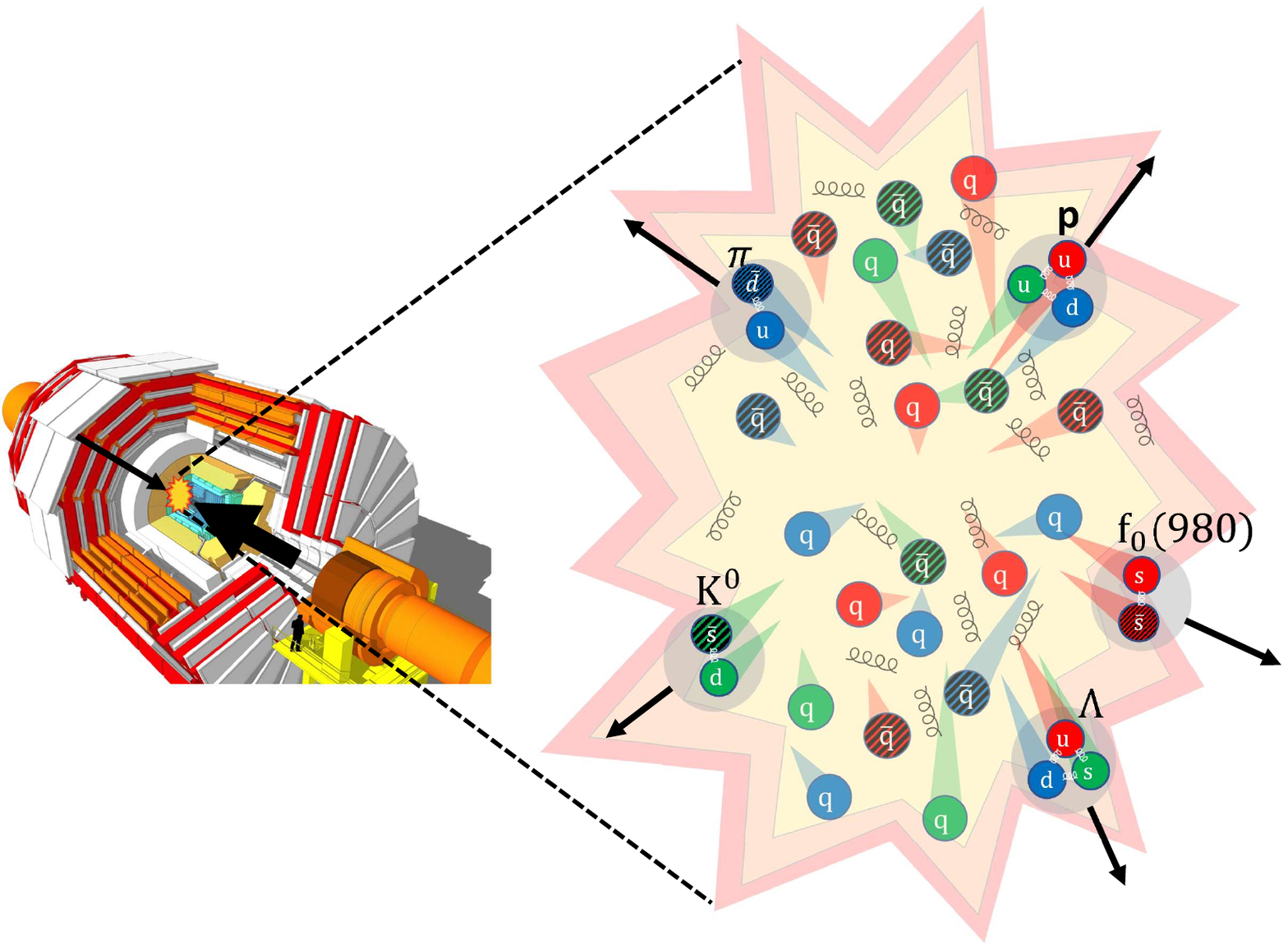}
    \caption{\textbf{Coalescence hadronization.} This picture illustrates the formation of hadrons in heavy ion collisions in the coalescence model. 
    Hadrons tend to form when the constituent quarks have similar positions and momenta. [Detector image reprinted from~\cite{Fig1}, under a \href{https://creativecommons.org/licenses/by-sa/4.0/}{CC BY SA 4.0 license}.]
    \label{fig:coalescence}}
\end{figure}

In the coalescence picture, illustrated in Fig.~\ref{fig:coalescence}, quarks with close spatial positions and momenta are 
more likely to combine and, therefore, the anisotropic flow coefficients \vn of the formed hadron inherit those of 
the parent quarks (\vnq). If \nq quarks with approximately equal momenta combine to form a hadron, the resulting azimuthal distribution is then given by 
\begin{equation}\begin{aligned}
\ifthenelse{\boolean{cms@external}}
{ 
\frac{\rd N_\Ph}{\rd\phi} &\propto\left(\frac{\rd N_\PQq}{\rd\phi}\right)^{\nq} \\
&\propto \left[1+\sum_{n=1}^{\infty}2\,\vnq(\pt^\PQq) \cos(n[\phi-\psi_{n}])\right]^{\nq}, 
\label{eq:flowh}
} 
{ 
\frac{\rd N_\Ph}{\rd\phi} &\propto\left(\frac{\rd N_\PQq}{\rd\phi}\right)^{\nq}\propto
\left[1+\sum_{n=1}^{\infty}2\,\vnq(\pt^\PQq) \cos(n[\phi-\psi_{n}])\right]^{\nq}, 
\label{eq:flowh}
} 
\end{aligned}\end{equation}
where $\pt^\PQq =\pt/\nq$, and $N_\Ph$ ($N_\PQq$) is the multiplicity of hadrons (quarks). 
For small values of \vn, relevant for the measurement reported in this paper, one can simplify Eq.~\eqref{eq:flowh} as
\begin{equation*}
\vn(\pt) \approx \nq \vnq(\pt/\nq).
\end{equation*}
This expression is commonly referred to as the number of constituent quarks (NCQ) scaling of the anisotropic flow~\cite{Molnar:2003ff}. 
The anisotropic flow of hadrons formed in heavy ion collisions can therefore reveal the number of constituent quarks \nq contained in a hadron, 
conventional or exotic~\cite{Gu:2019oyz}. Alternatively, such information can also be extracted by measuring the yields and \pt spectra 
(or their ratios) of these hadrons in heavy ion collisions, albeit in a more model-dependent way~\cite{Maiani:2006ia,ExHIC:2010gcb,Gu:2023whn,ALICE:2023cxn}.

The NCQ scaling has been observed to approximately hold for common hadrons in heavy ion collisions at the BNL RHIC~\cite{Adams:2003am, Adare:2006ti} 
and at the CERN LHC~\cite{ALICE:2014wao, ALICE:2018yph, CMS:2014und}. It has also been established in \ppb collisions at the LHC by the 
CMS experiment for \PKzS, \PGL, \PGXm, and \PGO hadrons~\cite{CMS:2014und,Sirunyan:2018toe}. 
These observations of the NCQ scaling support the validity of the coalescence hadronization model, at least at low \pt. 
We note that the NCQ scaling may additionally arise from other mechanisms in the same or an extended \pt range. 
The empirical observations of the NCQ scaling do not depend, however, on a particular underlying physics mechanism.

This paper presents the first measurement of the elliptic flow of the \fz state. Data from \ppb collisions at a nucleon-nucleon 
center-of-mass energy of $\sqrtsNN=8.16\TeV$ are used. The choice of \ppb collisions used in this measurement is driven by 
the smaller combinatorial background than in PbPb collisions, which simplifies the \fz signal extraction. 
The elliptic flow coefficient \vtwo of the \fz state is determined as a function of \pt. 
The NCQ scaling of the \fz hadron \vtwo coefficient is tested. We demonstrate that the hypothesis of the \fz state being an ordinary $\PQq\PAQq$ meson is significantly preferred over alternative hypotheses. 
This novel technique could be used to investigate other exotic hadron candidates. 
The numeric values from various figures presented in this paper can be found in the HEPData database~\cite{HEPData}.

\section{Results}
\subsection{Analysis of \texorpdfstring{\fz}{f0(980)} signal}
In this paper, the \fz state is measured in \ppb collisions at $\sqrtsNN=8.16\TeV$ by the CMS experiment. The CMS detector is described in Methods. A high-multiplicity data sample collected in 2016 is used, corresponding to an integrated luminosity of 186\nbinv~\cite{CMS:2016ngn}. The charged-particle multiplicity range is chosen to be $\nrange$. The \Ntrk multiplicity observable is defined in Ref.~\cite{Sirunyan:2018toe} and its range is chosen to be identical to the one used in that measurement, where significant anisotropic flow has been observed and to which we compare the NCQ scaling of the \fz measurement.
The triggers and event selections are identical to those in Ref.~\cite{CMS:2017kcs}, as discussed in more detail in the Methods section.

The \fz state is reconstructed within the rapidity $\abs{y}\lesssim 2.4$ 
via its dominant decay channel, $\fz \to \PGpp\PGpm$~\cite{ParticleDataGroup:2024cfk}. The pion mass is assigned to all charged-particle tracks. The combinatorial background is modeled via same-charge-sign pion track pairs 
and subtracted from the opposite-charge-sign dipion mass spectrum. The resulting distribution is then fit to extract the \fz yield. The fit model includes a sum of three Breit--Wigner functions~\cite{Weisskopf:1930au,Hull1959,STAR:2003vqj,Zhao:2022zix} corresponding to the \fz, 
$\PGrP{770}^0$, and $\PfDtP{1270}$ resonances, and a third-order polynomial for the residual background. Details of the fit procedure are described in the Methods section.

The observed elliptic flow \vtwo of the \fz state is extracted by fitting the yield as a function of $\phi$. 
The contamination from nonflow correlations---those unrelated to the nuclear collision geometry---is subtracted. After nonflow-contamination subtraction, the elliptic flow coefficient is denoted by \vsub. 
More details can be found in the Methods section.

In order to compare the \fz \vsub values to the established NCQ scaling for other hadrons, the \vnq of \PKzS, \PGL, \PGXm, and \PGO states measured in the same high-multiplicity range are fit with the following empirical function derived from data:
\ifthenelse{\boolean{cms@external}}
{ 
\begin{multline}
f(\etnq)\\
= \etnq  \left( p_0  + p_1 \etnq \right) \, \re^{-p_2 \etnq}. \label{equ:ncq}
\end{multline}
} 
{ 
\begin{equation}
f(\etnq)=\etnq  \left( p_0  + p_1 \etnq \right) \, \re^{-p_2 \etnq}. \label{equ:ncq}
\end{equation}
} 
The argument of the function, \etnq, is related to the kinetic energy per constituent quark, where $KE_\text{T}=\sqrt{\smash[b]{m^2+\ptmean^2}}-m$, \ptmean is the average \pt of a \pt bin of the corresponding bound state, 
and $m$ is its invariant mass. The \fz \ptmean values of the \pt bins are obtained from an exponential fit to the \fz candidate $\ddinline{N}{\pt}$ distribution. The $KE_\text{T}$ variable is chosen to describe the NCQ scaling as it yields better agreement with the data than \pt~\cite{Adare:2006ti}. The NCQ scaling fit is based on the minimization of the $\chi^2$, 
assuming that the bin-by-bin uncertainties are uncorrelated, with the coefficients $p_i$ ($i = 0$, 1, 2) being free parameters of the fit. Details about the \nq extraction and about testing of 
various quark content hypotheses can be found in the Methods section.

\subsection{Systematic uncertainties}
Systematic uncertainties in the \fz \vtwo and \vsub are detailed in the Methods section. The correlation of systematic uncertainties between different \pt bins is taken into account by using 
a covariance matrix in the $\chi^2$ calculation when extracting \nq. The statistical uncertainty in the $f(\etnq)$ fit is also included as a systematic uncertainty component in the extracted \nq. 
The \nq extraction procedure is repeated for variations in the functional form of $f(\etnq)$, as well as by using \pt instead of \ET in the NCQ scaling expression given by Eq.~\eqref{equ:ncq} (as discussed in the Methods section). 
The resulting difference in \nq from the default value is taken as the corresponding systematic uncertainty. The systematic uncertainties are listed in Table~\ref{tab:uncertainty_nq}. 
The uncertainty in the \ptmean of the \fz state has a negligible impact on \nq. 

\begin{table}[htb]
\centering
\topcaption{\label{tab:uncertainty_nq} Sources and magnitudes of the uncertainties in the extracted \nq of the \fz state in the range $\pt<10\GeVc$.}
    \begin{tabular}{l c}
\hline 
 Source & \nq uncertainty \\
\hline
Statistical                         & 0.16 \\
\fz \vtwo systematic uncertainty    & 0.13 \\
Nonflow effects in \vsub            & 0.04 \\
NCQ scaling fit parameters          & 0.02 \\
NCQ scaling fit function            & 0.04 \\
NCQ scaling using $\pt/\nq$         & 0.06 \\
\hline
\end{tabular}
\end{table}

\subsection{Elliptic anisotropy of \texorpdfstring{\fz}{f0(980)}}
\begin{figure*}[hbt]
\centering
\includegraphics[width=0.8\linewidth]{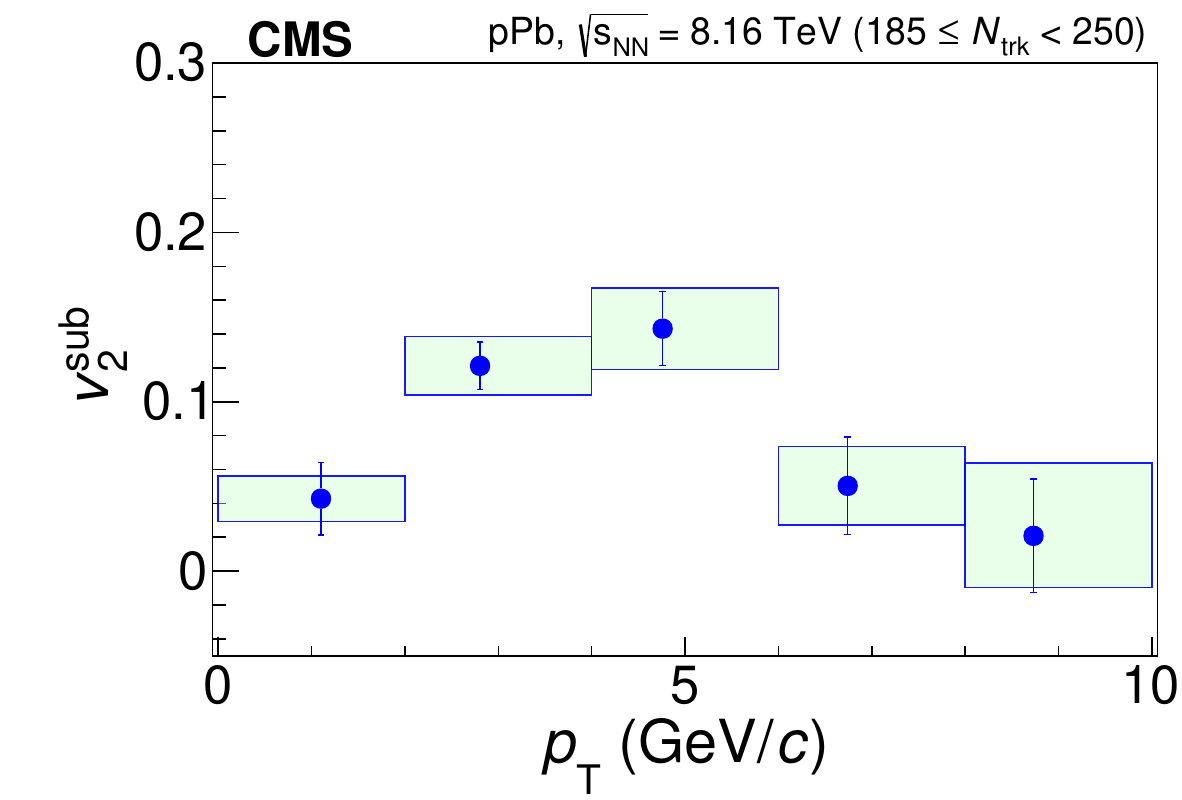}
\caption{\textbf{Elliptic anisotropy results.} The nonflow-effect-subtracted elliptic anisotropy \vsub of the \fz is shown as a function of \pt within $\abs{y} \lesssim 2.4$ in high-multiplicity \ppb collisions. 
The error bars show statistical uncertainties while the shaded areas represent systematic uncertainties.}
\label{fig:v2}
\end{figure*}

Figure~\ref{fig:v2} shows the \vsub of the \fz state, which is significantly above zero and exhibits a clear trend of rising and then falling with \pt, reaching a maximum in the $4<\pt<6\GeVc$ range. 
Such a trend in \pt has been also observed for other hadrons and is generally considered to come from an interplay between the hydrodynamic expansion at low \pt and partonic energy loss at high \pt.

\subsection{Quark content of \texorpdfstring{\fz}{f0(980)}}
\begin{figure}[hbt]
\centering
\includegraphics[width=\figonewidth]{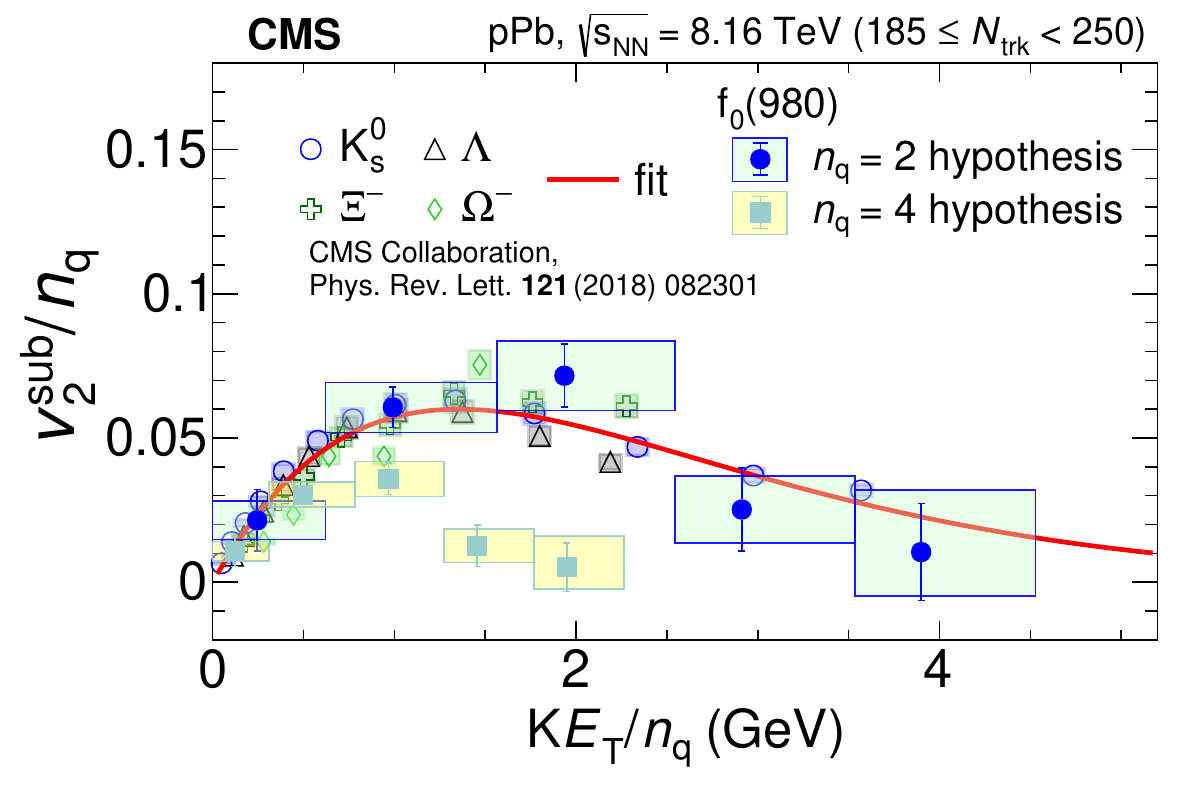}
\caption{\textbf{NCQ scaling of elliptic anisotropy.} The \vsubnq of the \fz state (for the $\nq=2$ and 4 hypotheses) as a function of \etnq, compared with those of \PKzS, \PGL, \PGXm, and \PGO 
strange hadrons~\cite{Sirunyan:2018toe} in high-multiplicity \ppb collisions. The error bars show statistical uncertainties while the shaded areas represent systematic uncertainties. 
The red curve is the NCQ scaling parameterization of the data for \PKzS, \PGL, \PGXm, and \PGO hadrons given by Eq.~\eqref{equ:ncq}.}
\label{fig:fig_mg_v2_DIVN} 
\end{figure}

Figure~\ref{fig:fig_mg_v2_DIVN} shows a comparison of \vsubnq for the \fz state with those of \PKzS, \PGL, \PGXm, and \PGO hadrons~\cite{Sirunyan:2018toe} as a function of \etnq. 
(A similar comparison for \vsubnq as a function of $\pt/\nq$ can be found in the Methods section.) The two sets of the \fz data points correspond to the $\nq=2$ and 4 hypotheses. 
The red curve shows the NCQ scaling parameterization of the \vsub data for these other hadrons (whose \nq values are fixed by their known quark content). 

\begin{figure}[hbt]
\centering
  \includegraphics[width=\figonewidth]{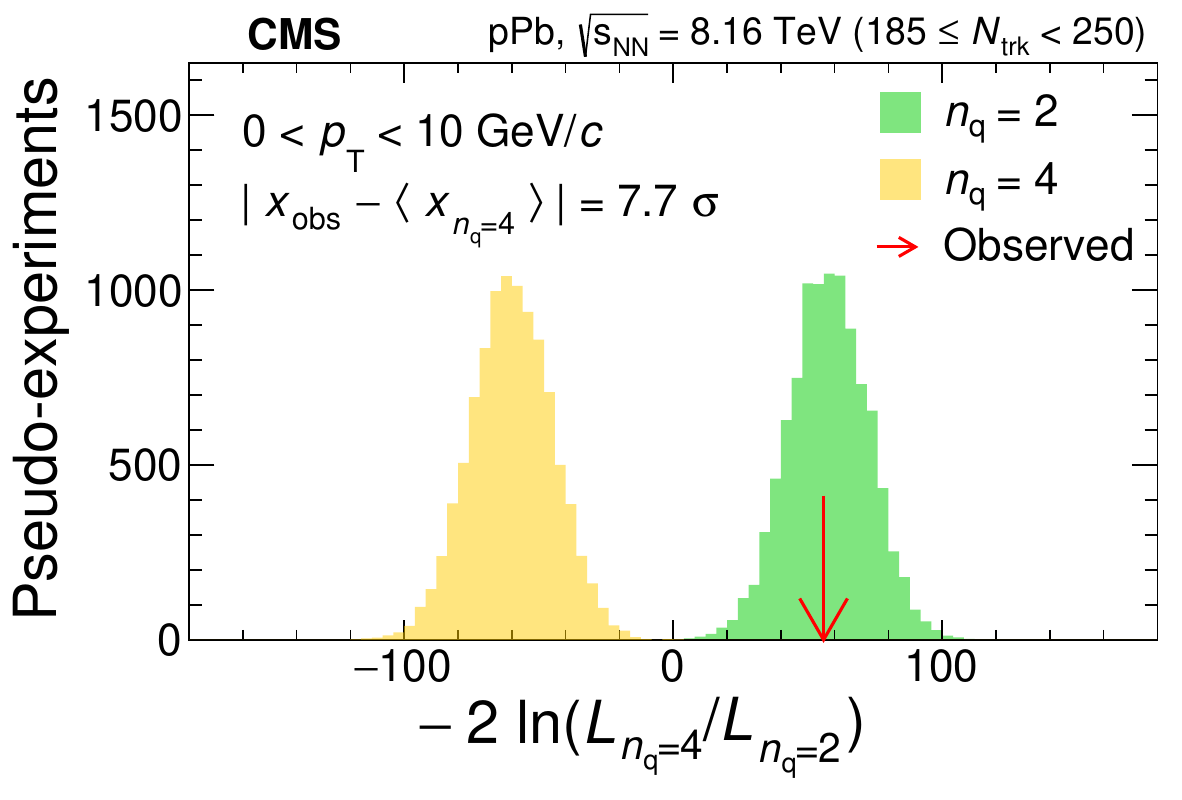}
  \caption{\textbf{Exclusion significance from $\nq=4$.} The log-likelihood ratio distributions for the $\nq=2$ and 4 hypotheses from pseudo-experiments, together with the measured value for the \fz state in the $0<\pt<10\GeVc$ range.}
  \label{fig:toymodel} 
\end{figure}

To assess the significance of the result, the log-likelihood ratio $-2\ln(L_{\nq=4}/L_{\nq=2})$ is calculated using the \vsubnq data and the NCQ scaling expectation between the $\nq=2$ and 4 assumptions. 
Details about the log-likelihood ratio can be found in the Methods section. The measured value is shown by the red arrow in Fig.~\ref{fig:toymodel}, together with the distributions 
of the log-likelihood ratio from pseudo-experiments. The \fz \vsub values are generated according to the NCQ scaling under the $\nq=2$ and 4 hypotheses, with a Gaussian smearing to account for the uncertainties. 
The extracted significance of the $\nq=2$ hypothesis over the $\nq=4$ hypothesis is 7.7 standard deviations ($\sigma$) in the $\pt<10\GeVc$ range. As shown in Fig.~\ref{fig:fig_mg_v2_DIVN}, the NCQ scaling range as delineated 
by the \PKzS data extends up to $\pt/\nq$ of $4\GeVc$, whereas for the baryons it is restricted to about $2.5\GeVc$. For the $\nq=2$ hypothesis, our high-\pt data start falling out of the 
measured NCQ scaling $\pt/\nq$ range; for the $\nq=4$ hypothesis, however, our data are within that range. Consequently, we extract significance values also for two restricted-\pt ranges: $\pt<8$ and $6\GeV/c$. 
The exclusion significances of the $\nq=4$ vs.\  2 hypotheses in these ranges are 6.3 and $3.1\sigma$, respectively.

The ${\PK\PAK}$ molecule, if produced by the coalescence of two kaons, would possess the same \vtwo as that of a tetraquark, and is thus practically also ruled out. 
It is unclear what \vtwo a hybrid $\PQq\PAQq\Pg$ state would attain in \ppb collisions because the NCQ scaling has been tested only with ordinary hadrons. 
If the constituent gluon behaves just like the constituent (anti)quarks, the \vtwo of a hybrid $\PQq\PAQq\Pg$ state would scale as $\nq=3$.
Such a state would be ruled out with a $3.5 \sigma$ significance using the $\pt<8\GeVc$ range, in which the NCQ scaling is adequately measured for the $\nq=3$ case.

The $\chi^2$ quantity is calculated between the \vnq data of the \fz, with floating \nq, and the NCQ curve in \etnq in Fig.~\ref{fig:fig_mg_v2_DIVN}, with the covariance matrix taking into account 
correlations among uncertainties. Scans of $\chi^2$ versus \nq are performed, as detailed in the Methods section. 
Using \fz data within the $\pt<6\GeVc$ range (a conservative choice, which ensures that the NCQ scaling holds for the $\nq =2$ hypothesis, given that $\pt/\nq < 3\GeVc$), the preferred \nq value 
of the \fz is found to be $\nq=2.40 \pm 0.40$. Assuming the NCQ scaling extends beyond $\pt/\nq \sim 3\GeVc$, the preferred \nq values of $2.10 \pm 0.24$ and $2.07 \pm 0.21$ are extracted 
in the $\pt<8$ and 10\GeVc ranges, respectively. Indeed, the $\nq=2$ hypothesis for \fz is consistent with the NCQ scaling from the other hadrons, with $\chi^2=4.7$ for the 5 data points.
Contrary to that, the $\nq=4$ hypothesis is inconsistent with the data, as evident from the corresponding $\chi^2=58$, with a Gaussian $p$-value of $3 \times 10^{-11}$. 
Consequently, we report a strong evidence for the $\PQq\PAQq$ quark content of the \fz state.

\section{Discussion}
The \fz state is observed in the $\PGpp\PGpm$ invariant mass distribution of high-multiplicity proton-lead collisions at $\sqrtsNN=8.16\TeV$, using  data collected by the CMS experiment in 2016 
and corresponding to an integrated luminosity of 186\nbinv. The elliptic flow anisotropy \vtwo of the \fz state is measured as a function of \pt up to 10\GeVc, with respect to the second-order 
harmonic plane reconstructed from forward/backward energy flow. After subtracting the nonflow contamination, evaluated from \PKzS measurements, we obtain the corrected \vsub observable. 
By comparing the \fz \vsub to those of \PKzS, \PGL, \PGXm, and \PGO under the number-of-constituent-quarks scaling hypothesis, we rule out the hypotheses that the \fz is a tetraquark state or a 
${\PK\PAK}$ molecule, in favor of an ordinary $\PQq\PAQq$ meson hypothesis, at 7.7$\sigma$ (6.3 or 3.1$\sigma$, respectively, if only a restricted range of $\pt<8$ or $6\GeVc$ is considered). 
The \fz data in the $\pt<8\GeVc$ range are found to disfavor a quark-antiquark-gluon hybrid state at 3.5$\sigma$. The number of constituent quarks of the \fz state, 
as extracted from a fit to the \vsub data, is consistent with the value of 2, characteristic of an ordinary meson. Consequently, we find strong evidence that the \fz hadron is a normal quark-antiquark state. 
We believe that the results reported in this paper offer a solution to a half-a-century-old puzzle. 

The experimental determination of the quark content of the \fz state with high confidence, using this novel approach, is expected to stimulate further experimental investigations as well as theoretical studies. 
It paves the way for studies of other exotic hadron candidates using the collective flow scaling approach in high-multiplicity proton-nucleus and heavy ion collisions.

\section{Methods}
In this section we provide experimental details of various steps used in the analysis presented in this paper.

\subsection{CMS detector}
The central feature of the CMS apparatus is a superconducting solenoid of 6\unit{m} internal diameter, providing a magnetic field of 3.8\unit{T}. Within the solenoid volume are a silicon pixel and strip tracker, a lead tungstate crystal electromagnetic calorimeter (ECAL), and a brass and scintillator hadron calorimeter (HCAL), each composed of a barrel and two endcap sections. Forward calorimeters extend the pseudorapidity coverage provided by the barrel and endcap detectors. Muons are measured in gas-ionization detectors embedded in the steel flux-return yoke outside the solenoid. The silicon tracker measures charged particles within the range $\abs{\eta} < 2.5$. For nonisolated particles of $1 < \pt < 10\GeVc$ and $\abs{\eta} < 1.4$, the track resolutions are typically 1.5\% in \pt and 25--90 (45--150)\mum in the transverse (longitudinal) impact parameter~\cite{CMS:2014pgm}. The procedure followed for aligning the detector is described in Ref.~\cite{CMS:2021ime}.

\subsection{Trigger}
Events of interest are selected using a two-tiered trigger system, a suite of triggers based on particle multiplicity. The first level (\mbox{level-1}), composed of custom hardware processors, uses information from the calorimeters and muon detectors to select events at a rate of around 100\unit{kHz} within a fixed latency of 4\mus~\cite{CMS:2020cmk}. At \mbox{level-1}, where tracking information is not available, the events were seeded using a tower count in the ECAL and HCAL barrel calorimeters, by selecting events passing a minimum threshold on the number of active towers. An active tower is defined as a trigger tower with a transverse energy exceeding 0.5\GeV. Trigger towers are built by summing energy deposits in the ECAL and HCAL cells in $\Delta\eta \times \Delta\phi = 0.087 \times 0.087$ regions (matching the size of one HCAL cell in the barrel region). The trigger required the tower count to exceed either $115$ or $120$, depending on the data-taking period.

The second level, known as the high-level trigger, consists of a farm of processors running a version of the full event reconstruction software optimized for fast processing, and reduces the event rate to around 1\unit{kHz} before data storage~\cite{CMS:2016ngn}. Several high-level triggers based on the multiplicity of tracks reconstructed either in the pixel detector layers or the full tracker were used for the analysis. The events were first selected requiring more than 125 tracks with $\pt > 0.4\GeVc$, $\abs{\eta}<2.4$, and the distance of closest approach along the beam axis between the track and the interaction vertex of less than 0.12\unit{cm}, reconstructed using only the pixel detector. Further, the events were required to have more than 185 tracks reconstructed in the full tracker with the same \pt and $\abs{\eta}$ requirements, and with the distance of closest approach less than 0.15\unit{cm}. The interaction vertex is required to be within 15\unit{cm} of the detector center along the beam direction. Offline, we require the number of reconstructed tracks, \Ntrk, to be between 185 and 250. The trigger turn-on effect has been shown to have a negligible impact on the result.

More detailed descriptions of the CMS detector, together with a definition of the coordinate system used and the relevant kinematic variables, can be found in Refs.~\cite{CMS:2008xjf,CMS:2023gfb}.

\subsection{Event selection and reconstruction of the \texorpdfstring{\fz}{f0(980)} signal}
\begin{figure*}[hbt]
\centering
\includegraphics[width=0.8\linewidth]{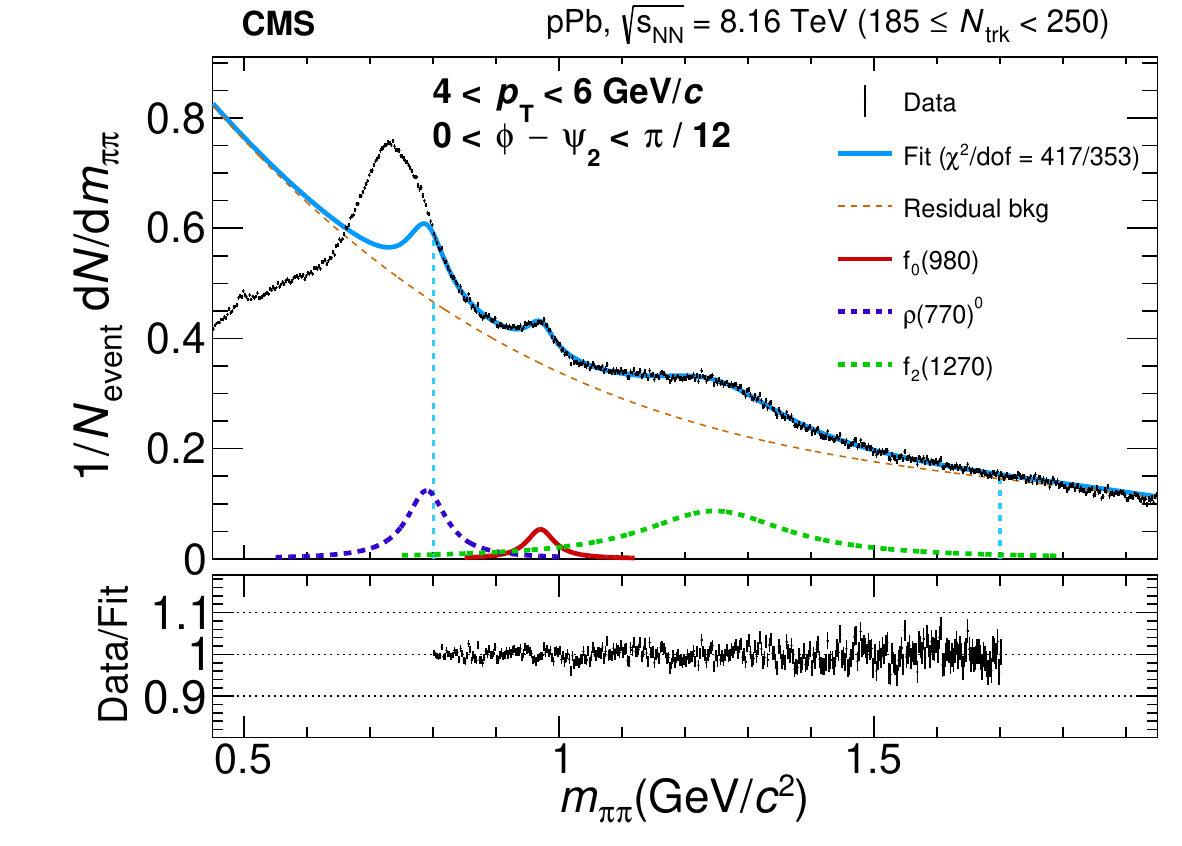}
\caption{\textbf{Invariant mass fit.} The invariant mass spectrum of opposite-sign pion pairs after the combinatorial background subtraction, for the pair transverse momentum $4<\pt<6\GeVc$ and 
the azimuthal angle $0<\dphi<\pi/12$, in high-multiplicity \ppb collisions. The solid blue curve is the fit result within the fit range marked with vertical blue dashed lines; 
the orange dashed curve represents the residual background. The solid red curve represents the \fz signal, while the dashed dark-violet and light-green curves correspond to the background contributions 
from the $\PGrP{770}^0$ and \Pfii resonances, respectively. The ratio between data and the fit result is shown in the lower panel, with the error bars representing statistical uncertainties only. 
The low-mass region exhibits a nontrivial turn-on behavior and is not included in the fit.
}\label{fig:mass_fit}
\end{figure*}

The \fz candidates are reconstructed through the dominant decay channel, $\fz \to \PGpp\PGpm$~\cite{ParticleDataGroup:2024cfk}. All charged-particle tracks with $\pt>0.4\GeVc$ and $\abs{\eta} < 2.4$ passing 
standard high-purity requirements~\cite{CMS:2014pgm}, and with a distance of closest approach to the interaction vertex divided by its uncertainty of less than 3 in both the direction along the beams 
and in the plane perpendicular to it, are considered as pion candidates. To improve the mass resolution, we only consider tracks with a relative uncertainty in \pt of less than 10\%. 
The \fz candidates are formed from pairs of tracks of opposite-charge-sign, with the charged pion mass assigned to both. The combinatorial background is estimated from same-charge-sign track pairs and is subtracted 
from the invariant mass spectrum of the \fz candidates. The spectrum is further corrected for the tracking efficiency as a function of the track \pt and $\eta$, as obtained via a \HIJING v1.0 
simulation~\cite{Gyulassy:1994ew} followed by the CMS detector response simulation with \GEANTfour~\cite{AGOSTINELLI2003250}. The analysis is performed in bins of \dphi, the azimuthal angle of the \fz candidate 
relative to that of the second harmonic plane. The latter is reconstructed from the energy deposition in the hadron forward (HF) calorimeter covering $3<\eta<5$ in the Pb-going direction (resulting in a better resolution compared to that using 
the opposite HF calorimeter) and corrected for the nonuniform detector performance by using the procedure described in Ref.~\cite{Poskanzer:1998yz}. 
Figure~\ref{fig:mass_fit} shows an example of the invariant mass spectrum of the \fz candidates within a \pt range of 4--6\GeVc and a \dphi range of 0--$\pi/12$ (where the \dphi value is first folded 
from the full range into the 0--$\pi/2$ range to decrease the statistical uncertainty per \dphi bin).

Several resonances are evident in the mass spectrum shown in Fig.~\ref{fig:mass_fit}, including a significant \fz peak at $\sim$0.98\GeVcc. 
The mass spectrum is fit with a template composed of three Breit--Wigner functions corresponding to the $\PGrP{770}^0$, \fz, and \Pfii resonances, and a third-order polynomial to model the residual background. 
The fit mass range is chosen to be 0.8--1.7\GeVcc in order to exclude the contribution from a $\PGrP{1700}$ peak at high masses and to avoid the low-mass region ($<$0.8\GeVcc) exhibiting a nontrivial turn-on behavior. 
Since only the right tail of the $\PGrP{770}^0$ resonance is included in the fit, the extrapolated peak into the lower mass region does not necessarily represent the true shape of the $\PGrP{770}^0$ resonance. 
The $\phi$-integrated mass spectrum is fit in each \pt interval to obtain the \fz yield and the line-shapes of the three resonances present within the fit window. The resonant line-shapes are then fixed and 
the fit of the mass spectrum is repeated in six individual \dphi bins in the corresponding \pt interval, treating the resonance yields as free parameters. The resultant fit to the example \dphi bin of the \pt interval is superimposed in Fig.~\ref{fig:mass_fit}, along with the $\chi^2$ of the fit per degree of freedom (dof).

\subsection{Extraction of \texorpdfstring{\fz}{f0(980)} elliptic anisotropy \texorpdfstring{$\mathbf\vtwo$}{v2} values}
\begin{figure}[hbtp]
\centering
\includegraphics[width=\figonewidth]{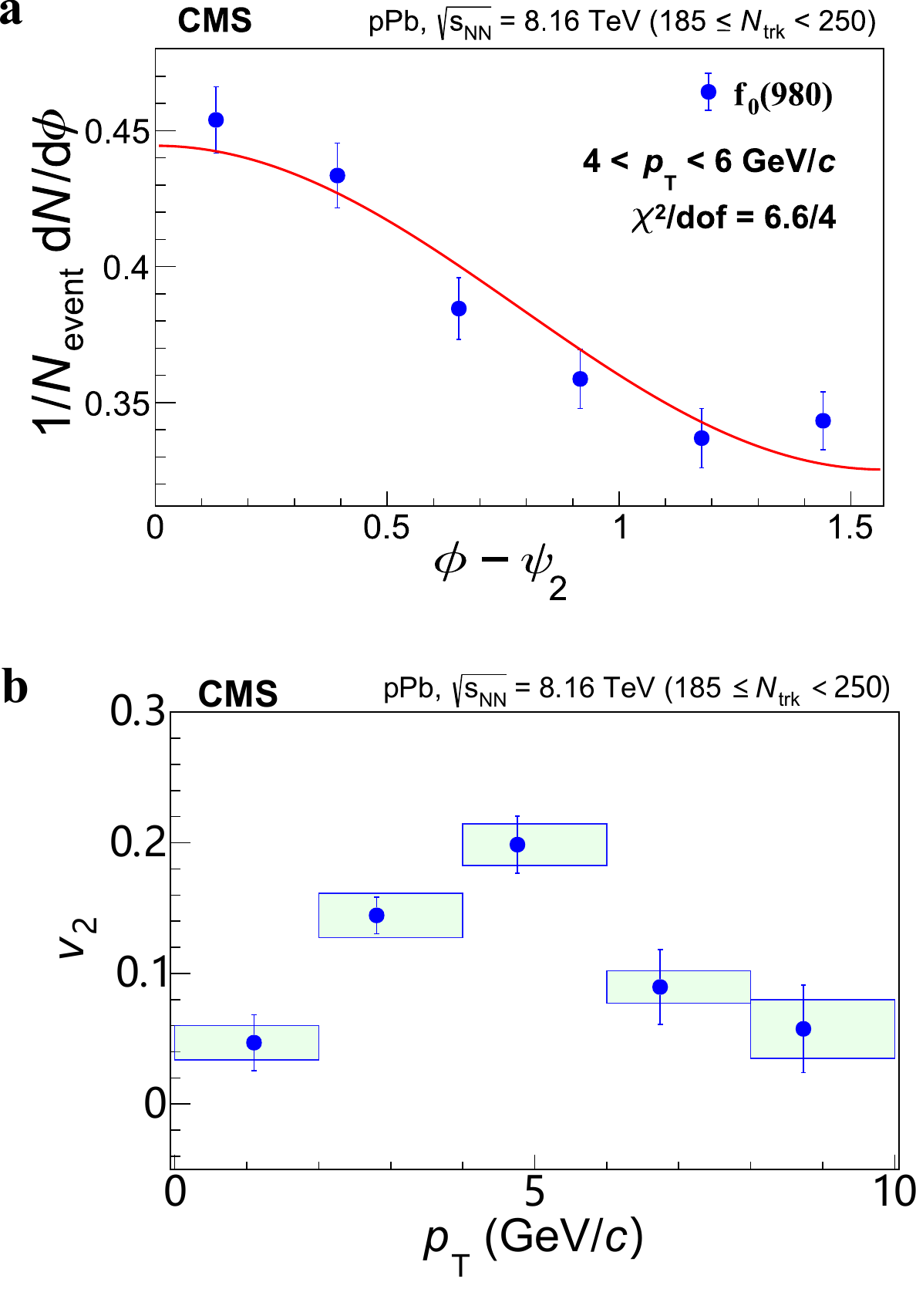}
\caption{\textbf{Elliptic anisotropy before the nonflow effect subtraction.} {\bf{a}}: The \fz yield in the $4<\pt<6\GeVc$ range as a function of \dphi in high-multiplicity \ppb collisions. 
Error bars show statistical uncertainties. The red curve is a fit to Eq.~\eqref{equ:flows} with only the $n=2$ term, from which the elliptic anisotropy \vtwo parameter is extracted. {\bf{b}}: The elliptic anisotropy \vtwo of the \fz state is shown before the nonflow effect subtraction as a function of \pt within rapidity $\abs{y} \lesssim 2.4$ 
in high-multiplicity \ppb collisions. The error bars show statistical uncertainties while the shaded areas represent systematic uncertainties.}\label{fig:yield_phi_fit_v2bef}
\end{figure}

Figure~\ref{fig:yield_phi_fit_v2bef}a shows the \fz yield as a function of \dphi in the $4<\pt<6\GeVc$ bin as an example. The \fz yield as a function of \dphi is fit with Eq.~\eqref{equ:flows} with only the $n=2$ term 
to extract the \vtwo parameter. The fitted \vtwo values are corrected for the harmonic plane resolution, which represents the precision of the reconstructed $\psi_2$. The resolution is obtained by the three-subevent method~\cite{Poskanzer:1998yz} and evaluated in each fine multiplicity interval, and an average resolution is obtained weighted by the corresponding $\phi$-integrated yield of \fz. 
The three-subevent method uses the two HFs and the central tracker detector, where the $\eta$ gaps between the subevents help suppressing the nonflow effects. 
Figure~\ref{fig:yield_phi_fit_v2bef}b shows the corrected \vtwo of the \fz as a function of \pt.

The \vtwo measurement is contaminated by nonflow correlations, such as back-to-back jet pairs, where an \fz candidate is found within a jet and the harmonic plane is reconstructed from hadrons that include 
the other fragments of the dijet system. Since \fz is a hadron known to likely contain strange quarks, the relative nonflow contribution $(\vtwo-\vsub)/\vtwo$ to the \fz \vtwo is assumed to be the same 
as that for the \PKzS meson, in each \pt bin, where \vsub represents the elliptic flow after nonflow-effect subtraction. The latter, evaluated using events with low track multiplicity~\cite{Sirunyan:2018toe}, 
is fit with a second-order polynomial as a function of \pt. The relative nonflow contribution to the \fz \vtwo is evaluated from the fit function at the \ptmean in each \pt bin and 
ranges 9--64\% for different \pt bins. The nonflow effects are subtracted to obtain the final \vsub of the \fz state. The \vsub of the \fz is shown in Fig.~\ref{fig:v2} as a function of \pt.

\subsection{Systematic uncertainties in \texorpdfstring{\fz}{f0(980)} \texorpdfstring{$\mathbf{\vtwo}$}{v2} values}
The systematic uncertainties in the \fz yield, and consequently those in the \fz \vtwo values, include those from track selection, track efficiency correction, combinatorial background subtraction, 
residual background parameterization, resonance line-shape modeling, fit range choice, and the harmonic plane resolution. They are summarized in Table~\ref{tab:uncertainty_nq} and are evaluated as follows. 
\begin{itemize}
\item Looser and tighter criteria of track selection are applied and the obtained \fz \vtwo results are compared to the default ones, all of which are not corrected for track efficiencies. 
The uncertainty is 3--22\% in the \fz \vtwo value, depending on \pt. 

\item There could be a difference in the detector response between the same-charge-sign and opposite-charge-sign pairs in the same event. To assess this systematic uncertainty, 
the default combinatorial background spectrum from same-sign pairs is scaled by the ratio of opposite- to same-sign spectra from mixed events (\ie, when the two tracks forming a pair are taken from different events). 
The effect on the \fz \vtwo values is smaller than 3\%.

\item The residual background is parameterized by second-, fourth-, and fifth-order polynomials besides the default third-order one. The corresponding systematic uncertainty in \vtwo is found to be less than 5\%.

\item The resonance mass peaks are alternatively modeled via a relativistic Breit--Wigner function~\cite{Matthews:1959sy} and a relativistic Voigt function~\cite{Kycia:2017gjn}, which yields 
a systematic uncertainty in \vtwo of less than 2\%, except in the highest measured \pt interval, where it reaches 25\%.
The default fit range (0.8--1.7\GeVcc) is varied by 0.02\GeVcc on each side and gives a systematic uncertainty in \vtwo of less than 8\%. 

\item The statistical uncertainty in the harmonic plane angle extraction is propagated to the \fz \vtwo and treated as a systematic uncertainty, of approximately 6\%.

\item An alternative way to estimate the nonflow contribution to the \fz \vtwo is by assuming that the absolute nonflow contribution $\vtwo-\vsub$, instead of the relative $(\vtwo-\vsub)/\vtwo$ contribution, 
at a given \pt, is the same as that of the \PKzS meson. The difference between the \fz \vsub estimates obtained with the alternative and default methods is treated as the systematic uncertainty 
in the nonflow-effect subtraction, which is further symmetrized using the larger of the negative and positive variations. The resultant systematic uncertainty band in \vsub is then capped between 
the measured \vtwo value and zero, ranging from 1\% to 33\%, depending on \pt.

\item We have also examined the nonflow contribution using \PDz and \PGL data instead of \PKzS data. Moreover, we have compared the uncorrected \vtwo distribution of the \fz to those of 
the \PKzS, \PGL, \PGXm, and \PGO hadrons. The variations observed in these cross-checks are shown to be fully covered by the systematic uncertainty detailed in the previous item.
\end{itemize}

These various sources of systematic uncertainties are assumed to be independent of each other. The statistical uncertainty is treated as uncorrelated for different \pt bins. 
The systematic uncertainty arising from the event plane resolution is assumed to be fully correlated between the \pt bins. For each of the other systematic uncertainties, the \vsub covariance matrix element 
for the $i$th and $j$th \pt bins is calculated as 
\begin{equation}\begin{aligned}
\ifthenelse{\boolean{cms@external}}
{ 
\sigma_{i,j}= \frac{1}{N_{\text{alt}}} & \sum_{k = 1}^{N_{\text{alt}}}  { \left( v_{2,k}^{\text{sub}}(p_{\mathrm{T},i})-v_{2,\text{default}}^{\text{sub}}(p_{\mathrm{T},i})\right)} \\
  &\times \left( v_{2,k}^{\text{sub}}(p_{\mathrm{T},j})-v_{2,\text{default}}^{\text{sub}}(p_{\mathrm{T},j}) \right), 
\label{eq:cov}
} 
{ 
\sigma_{i,j} &= \frac{1}{N_{\text{alt}}} \sum_{k = 1}^{N_{\text{alt}}} { \left( v_{2,k}^{\text{sub}}(p_{\mathrm{T},i})-v_{2,\text{default}}^{\text{sub}}(p_{\mathrm{T},i})\right) \left(v_{2,k}^{\text{sub}}(p_{\mathrm{T},j})-v_{2,\text{default}}^{\text{sub}}(p_{\mathrm{T},j}) \right)}, \label{eq:cov}
} 
\end{aligned}\end{equation}
where $N_{\text{alt}}$ is the number of alternative methods to extract this systematic uncertainty and $v_{2,\text{default}}^{\text{sub}}$ is the default value. 
The overall covariance matrix of the \vsub is the sum of the covariance matrices corresponding to the various systematic uncertainties. 
The uncertainty in the \ptmean evaluation is estimated using pseudo-experiments and is found to be negligible.

\subsection{Cross-checks of the NCQ scaling assumption}
The uncertainty used in the parametrization of Eq.~\eqref{equ:ncq} of the \vsubnq data for the \PKzS, \PGL, \PGXm, and \PGO hadrons is the combined statistical and systematic uncertainties of the fit added in quadrature. 
The resulting best fit parameters are: $p_0=0.111\pm 0.004$, $p_1=0.045 \pm 0.017$, and $p_2=-1.00 \pm 0.08$. The fit yields a relatively large $\chi^2/{\text{dof}}=80/34$, which indicates that 
the NCQ scaling is not perfect. To accommodate for this, we increased the uncertainties to achieve the $\chi^2/{\text{dof}}$ of 1 and found the effect on the \nq result to be negligible.

\begin{figure}[hbtp]
\centering
\includegraphics[width=\figonewidth]{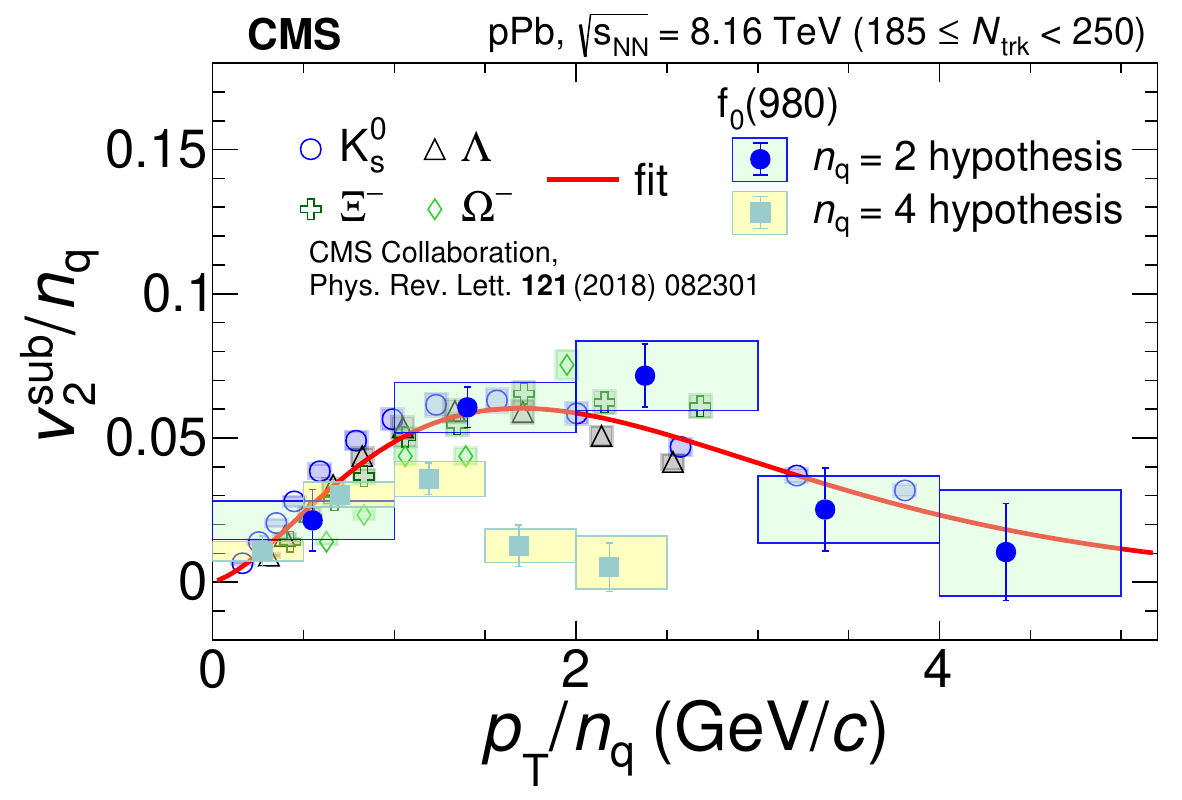}
\caption{\textbf{NCQ scaling of elliptic anisotropy in $\pt/\nq$.} The \vsubnq of the \fz state (for the $\nq=2$ and 4 hypotheses) as a function of $\pt/\nq$ is compared 
with those of the \PKzS, \PGL, \PGXm, and \PGO strange hadrons~\protect\cite{Sirunyan:2018toe} in high-multiplicity \ppb collisions. Error bars show the statistical uncertainties while the shaded areas 
represent systematic uncertainties. The red curve is the NCQ scaling parameterization of the data for the \PKzS, \PGL, \PGXm, and \PGO hadrons.}
\label{fig:v2nqpt}
\end{figure}

The NCQ scaling can also be parameterized in $\pt/\nq$. The fit quality is worse, with $\chi^2/{\text{dof}}=170/34$. The NCQ-scaled \vsubnq as a function of $\pt/\nq$ is shown in Fig.~\ref{fig:v2nqpt} for the \fz state together with those of the \PKzS, \PGL, \PGXm, and \PGO hadrons~\cite{Sirunyan:2018toe}. Using NCQ scaling in $\pt/\nq$, the extracted significance of the $\nq =2$ hypothesis over the $\nq =4$ hypothesis is 7.8, 6.2, or $2.4\sigma$, in the $\pt <10$, 8, or 6\GeVc ranges, respectively.

\begin{figure}[hbtp]
\centering
  \includegraphics[width=\figonewidth]{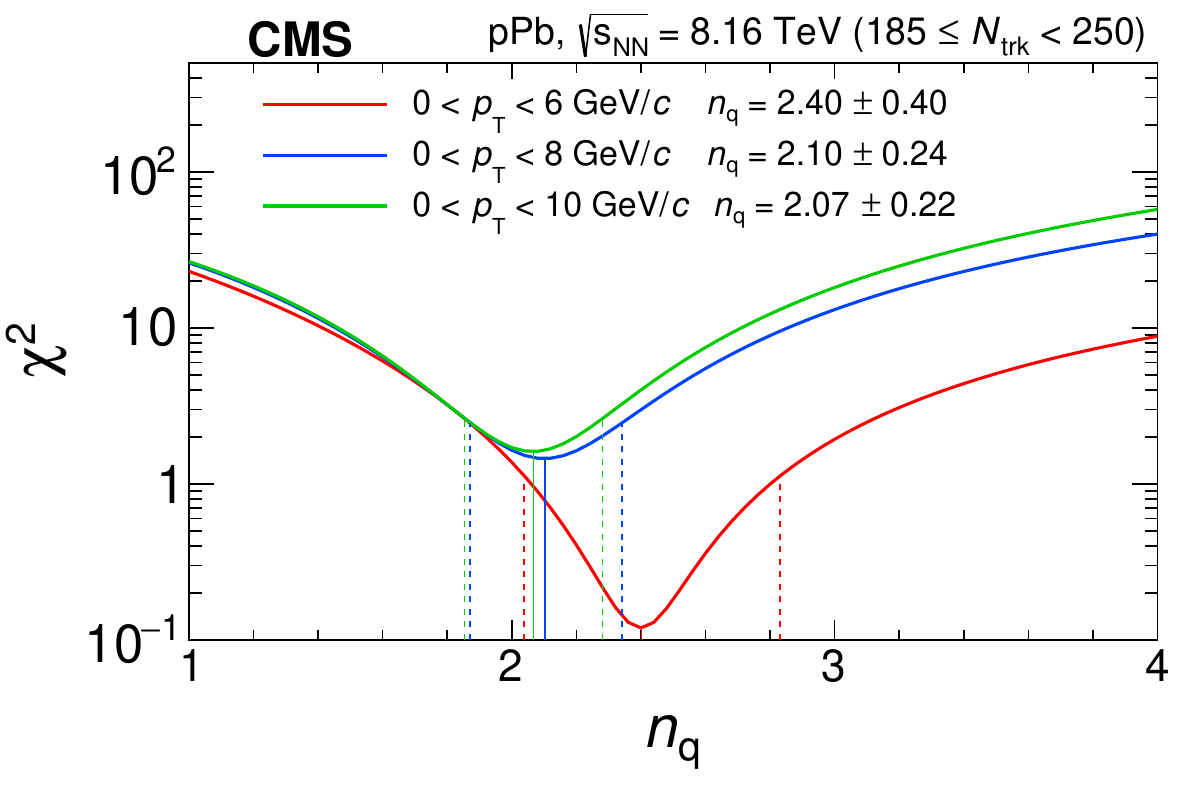}
  \caption{\textbf{The $\chi^2$ scan.} The $\chi^2$ of the \fz elliptic flow data with respect to the NCQ scaling parameterization, scanned in steps of $n_\PQq$.
  The three curves correspond to using \fz data for $\pt<6$, 8, and 10\GeVc, respectively.}
 \label{fig:chi2_nq} 
\end{figure}

To extract the \nq of \fz, the $\chi^2$ of the \fz \vsubnq data (denoted by $\vec{y}$) with respect to the NCQ scaling curve (denoted by $\vec{f}$) is calculated by 
$\chi^2 = (\vec{y} - \vec{f})^T  (C_y + C_f)^{-1}   (\vec{y} - \vec{f})$, where $C_y$ is the \fz \vsub covariance matrix scaled by $1/\nq^2$ and $C_f$ is the covariance matrix between the NCQ scaling function 
from the fit and the \fz \vsubnq data. The latter is given by $C_f=J \cdot C_p \cdot J^T$, where $C_p$ is the covariance matrix of the fit parameters and $J=\partial f/\partial \vec{p}$ is 
the Jacobian matrix that describes how the fit function value changes with the fit parameters $\vec{p}$. 

Scans of $\chi^2$ as a function of \nq, treated as a continuous parameter, between the \fz \vsubnq data and the NCQ scaling curve, are shown in Fig.~\ref{fig:chi2_nq}. 
The three curves correspond to the \fz data for $\pt<6$, 8, and 10\GeVc, respectively. The statistical and systematic uncertainties are included in the $\chi^2$ calculation with the covariance matrix. 
The optimal \nq value is determined at the minimum $\chi^2$, denoted by $\chi_\text{min}^2$, with the uncertainty bracketed by the $\chi_{\rm min}^2+1$ level. 
The corresponding \nq values are listed in Fig.~\ref{fig:chi2_nq}, with the uncertainties accounting for the effects of variations in the NCQ fit functional forms and from using $\pt/\nq$ instead of \etnq, 
which are relatively small (as detailed in Table~\ref{tab:uncertainty_nq}).

Lead-lead collision data from ALICE suggest that the NCQ scaling holds within a precision of 20\%~\cite{ALICE:2014wao, ALICE:2018yph}. 
We vary the overall fit curve of the NCQ scaling by $\pm$10\% and obtain $\nq=1.95 \pm 0.22$ and $2.19 \pm 0.24$, respectively for the positive and negative variations, by using 
\fz data for $\pt<10\GeVc$. Both these values agree well with the nominal result of $\nq=2.07 \pm 0.22$, demonstrating the robustness of our result with respect to variations in the NCQ scaling assumption. 

The NCQ scaling may be less valid at low \pt where \vsub likely reflects hydrodynamic behavior which is mass dependent. However, excluding the lowest \pt data point has negligible effect on our results. 

To examine the effects of imperfect NCQ scaling, we carry out further cross-checks as follows. We fit only the \PKzS \vsub data to obtain an alternative NCQ scaling curve and repeat the analysis. 
The \fz \nq extracted this way is $2.03 \pm 0.22$. When we use the nominal NCQ scaling curve with \PGL \vsub data, the extracted \PGL \nq value is $2.73 \pm 0.14$. 
The $2\sigma$ deviation from the nominal value of 3 indicates degree of the validity of the NCQ scaling between the \PKzS and \PGL hadrons. Similarly, when we use only the \PGL \vsub data to obtain the NCQ scaling curve, 
the extracted \nq is $2.30 \pm 0.22$ for the \fz and $2.29 \pm 0.18$ for the \PKzS states. We have also tested the NCQ scaling validity on $\PGO$ data by using the \PKzS, \PGL, and \PGXm data 
for the NCQ scaling fit; the extracted \nq value for \PGO is $3.21 \pm 0.69$.

\subsection{Exclusion significance determination}
The log-likelihood ratio $-2\ln(L_{\nq=4}/L_{\nq=2})$, evaluated as the $\chi^2$ difference between the \mbox{$\nq=2$} and 4 hypotheses, is calculated for the \fz \vsub data. 
We also generate pseudo-data of \fz \vsub according to the NCQ scaling curve for a given \nq hypothesis. The \vsub uncertainties are taken into account by smearing \vsub with a Gaussian distribution 
according to the covariance matrix given by Eq~\eqref{eq:cov}. The corresponding log-likelihood ratio is calculated for the pseudo-data in the same way as for \ppb data. 
The pseudo-experiments yield an expected distribution of the log-likelihood ratio for the two given \fz \nq hypotheses. Each of these distributions is fit with a Gaussian function and the significance 
of the main hypothesis over the alternative one is extracted as the distance between the Gaussian mean of the alternative distribution (the yellow histogram in, \eg, Fig.~\ref{fig:toymodel}) and 
the measured value in data (the red arrow in the same figure), divided by the width of the Gaussian function. The consistency with the main hypothesis can be inferred in a similar way 
by comparing the value obtained in data with the Gaussian mean and width of the main distribution (the green histogram in the same figure).

\begin{figure}[hbtp]
\centering
  \includegraphics[width=\figonewidth]{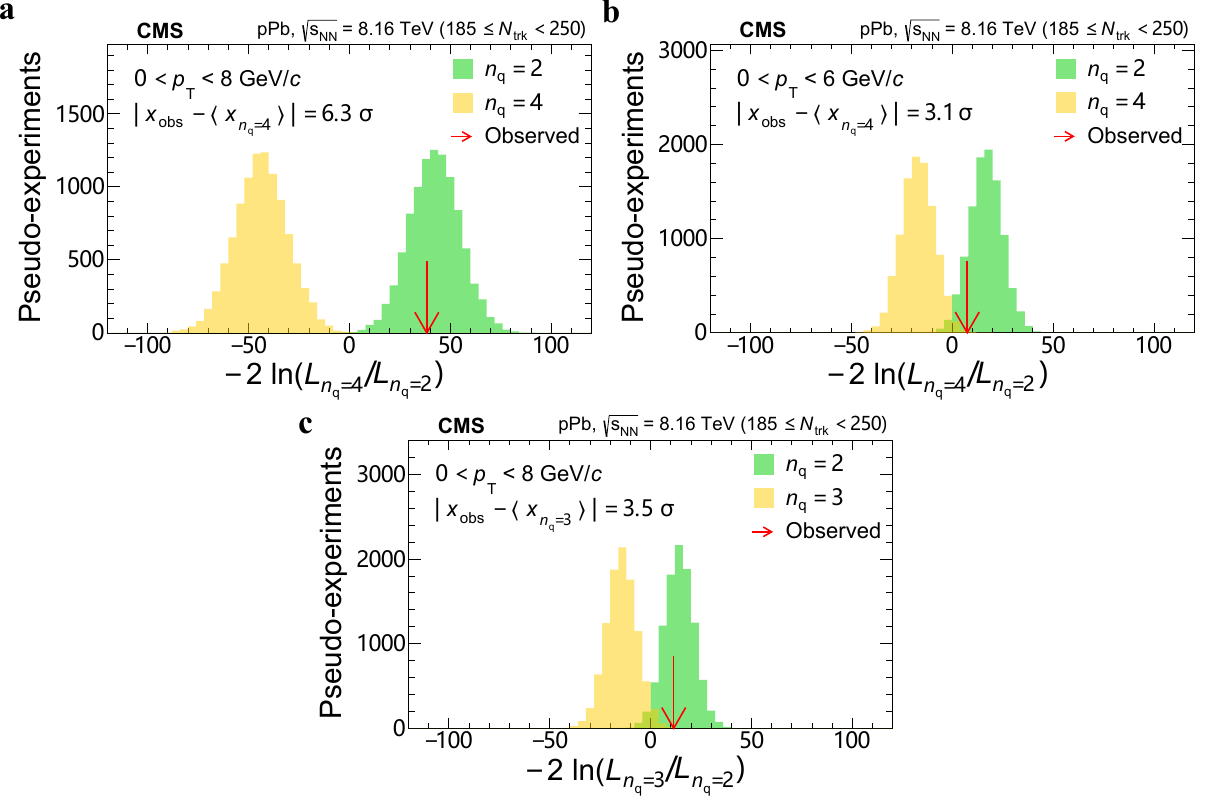}
  \caption{\textbf{Exclusion significances.} {\bf{a}}, {\bf{b}}: same as Fig.~\ref{fig:toymodel} but using \fz \vsub data within the restricted ranges $\pt<8\GeVc$ ({\bf{a}}) and $\pt<6\GeVc$ ({\bf{b}}). {\bf{c}}: The expected log-likelihood ratio distributions for $\nq=2$ vs.\ 3 hypotheses from the pseudo-experiments and the observed value for the \fz in data in the $\pt<8\GeVc$ range to extract the exclusion significance for $\nq=3$.}
  \label{fig:toy} 
\end{figure}

The log-likelihood ratio distributions for the $\nq =2$ vs.\ 4 hypotheses for the two restricted \pt ranges are shown in Fig.~\ref{fig:toy}a,~\ref{fig:toy}b. The log-likelihood ratio distributions for 
the $\nq=2$ vs.\ 3 hypotheses in the $\pt < 8\GeVc$ range are shown in Fig.~\ref{fig:toy}c.

\begin{acknowledgments}
  We congratulate our colleagues in the CERN accelerator departments for the excellent performance of the LHC and thank the technical and administrative staffs at CERN and at other CMS institutes for their contributions to the success of the CMS effort. In addition, we gratefully acknowledge the computing centers and personnel of the Worldwide LHC Computing Grid and other centers for delivering so effectively the computing infrastructure essential to our analyses. Finally, we acknowledge the enduring support for the construction and operation of the LHC, the CMS detector, and the supporting computing infrastructure provided by the following funding agencies: SC (Armenia), BMBWF and FWF (Austria); FNRS and FWO (Belgium); CNPq, CAPES, FAPERJ, FAPERGS, and FAPESP (Brazil); MES and BNSF (Bulgaria); CERN; CAS, MoST, and NSFC (China); MINCIENCIAS (Colombia); MSES and CSF (Croatia); RIF (Cyprus); SENESCYT (Ecuador); MoER, ERC PUT and ERDF (Estonia); Academy of Finland, MEC, and HIP (Finland); CEA and CNRS/IN2P3 (France); SRNSF (Georgia); BMBF, DFG, and HGF (Germany); GSRI (Greece); NKFIH (Hungary); DAE and DST (India); IPM (Iran); SFI (Ireland); INFN (Italy); MSIP and NRF (Republic of Korea); MES (Latvia); LAS (Lithuania); MOE and UM (Malaysia); BUAP, CINVESTAV, CONACYT, LNS, SEP, and UASLP-FAI (Mexico); MOS (Montenegro); MBIE (New Zealand); PAEC (Pakistan); MES and NSC (Poland); FCT (Portugal); MESTD (Serbia); MCIN/AEI and PCTI (Spain); MOSTR (Sri Lanka); Swiss Funding Agencies (Switzerland); MST (Taipei); MHESI and NSTDA (Thailand); TUBITAK and TENMAK (Turkey); NASU (Ukraine); STFC (United Kingdom); DOE and NSF (USA).
 
  \hyphenation{Rachada-pisek} Individuals have received support from the Marie-Curie program and the European Research Council and Horizon 2020 Grant, contract Nos.\ 675440, 724704, 752730, 758316, 765710, 824093, and COST Action CA16108 (European Union); the Leventis Foundation; the Alfred P.\ Sloan Foundation; the Alexander von Humboldt Foundation; the Science Committee, project no. 22rl-037 (Armenia); the Belgian Federal Science Policy Office; the Fonds pour la Formation \`a la Recherche dans l'Industrie et dans l'Agriculture (FRIA-Belgium); the Agentschap voor Innovatie door Wetenschap en Technologie (IWT-Belgium); the F.R.S.-FNRS and FWO (Belgium) under the ``Excellence of Science -- EOS" -- be.h project n.\ 30820817; the Beijing Municipal Science \& Technology Commission, No. Z191100007219010 and Fundamental Research Funds for the Central Universities (China); the Ministry of Education, Youth and Sports (MEYS) of the Czech Republic; the Shota Rustaveli National Science Foundation, grant FR-22-985 (Georgia); the Deutsche Forschungsgemeinschaft (DFG), under Germany's Excellence Strategy -- EXC 2121 ``Quantum Universe" -- 390833306, and under project number 400140256 - GRK2497; the Hellenic Foundation for Research and Innovation (HFRI), Project Number 2288 (Greece); the Hungarian Academy of Sciences, the New National Excellence Program - \'UNKP, the NKFIH research grants K 124845, K 124850, K 128713, K 128786, K 129058, K 131991, K 133046, K 138136, K 143460, K 143477, 2020-2.2.1-ED-2021-00181, and TKP2021-NKTA-64 (Hungary); the Council of Science and Industrial Research, India; ICSC -- National Research Center for High Performance Computing, Big Data and Quantum Computing, funded by the EU NexGeneration program (Italy); the Latvian Council of Science; the Ministry of Education and Science, project no. 2022/WK/14, and the National Science Center, contracts Opus 2021/41/B/ST2/01369 and 2021/43/B/ST2/01552 (Poland); the Funda\c{c}\~ao para a Ci\^encia e a Tecnologia, grant CEECIND/01334/2018 (Portugal); the National Priorities Research Program by Qatar National Research Fund; MCIN/AEI/10.13039/501100011033, ERDF ``a way of making Europe", and the Programa Estatal de Fomento de la Investigaci{\'o}n Cient{\'i}fica y T{\'e}cnica de Excelencia Mar\'{\i}a de Maeztu, grant MDM-2017-0765 and Programa Severo Ochoa del Principado de Asturias (Spain); the Chulalongkorn Academic into Its 2nd Century Project Advancement Project, and the National Science, Research and Innovation Fund via the Program Management Unit for Human Resources \& Institutional Development, Research and Innovation, grant B37G660013 (Thailand); the Kavli Foundation; the Nvidia Corporation; the SuperMicro Corporation; the Welch Foundation, contract C-1845; and the Weston Havens Foundation (USA).  
\end{acknowledgments}

\bibliography{auto_generated}

\cleardoublepage \section{The CMS Collaboration \label{app:collab}}\begin{sloppypar}\hyphenpenalty=5000\widowpenalty=500\clubpenalty=5000
\cmsinstitute{Yerevan Physics Institute, Yerevan, Armenia}
{\tolerance=6000
A.~Hayrapetyan, A.~Tumasyan\cmsAuthorMark{1}\cmsorcid{0009-0000-0684-6742}
\par}
\cmsinstitute{Institut f\"{u}r Hochenergiephysik, Vienna, Austria}
{\tolerance=6000
W.~Adam\cmsorcid{0000-0001-9099-4341}, J.W.~Andrejkovic, T.~Bergauer\cmsorcid{0000-0002-5786-0293}, S.~Chatterjee\cmsorcid{0000-0003-2660-0349}, K.~Damanakis\cmsorcid{0000-0001-5389-2872}, M.~Dragicevic\cmsorcid{0000-0003-1967-6783}, P.S.~Hussain\cmsorcid{0000-0002-4825-5278}, M.~Jeitler\cmsAuthorMark{2}\cmsorcid{0000-0002-5141-9560}, N.~Krammer\cmsorcid{0000-0002-0548-0985}, A.~Li\cmsorcid{0000-0002-4547-116X}, D.~Liko\cmsorcid{0000-0002-3380-473X}, I.~Mikulec\cmsorcid{0000-0003-0385-2746}, J.~Schieck\cmsAuthorMark{2}\cmsorcid{0000-0002-1058-8093}, R.~Sch\"{o}fbeck\cmsorcid{0000-0002-2332-8784}, D.~Schwarz\cmsorcid{0000-0002-3821-7331}, M.~Sonawane\cmsorcid{0000-0003-0510-7010}, S.~Templ\cmsorcid{0000-0003-3137-5692}, W.~Waltenberger\cmsorcid{0000-0002-6215-7228}, C.-E.~Wulz\cmsAuthorMark{2}\cmsorcid{0000-0001-9226-5812}
% ...
\par}
\vskip\cmsinstskip
\dag:~Deceased\\
$^{1}$Also at Yerevan State University, Yerevan, Armenia\\
$^{2}$Also at TU Wien, Vienna, Austria\\
% ...
\end{sloppypar}
\end{document}